\documentclass[aps,prmaterials,reprint,superscriptaddress,amsmath,amssymb,nofootinbib]{revtex4-2}

\usepackage[T1]{fontenc}
\usepackage{graphicx}
\usepackage{hyperref}
\hypersetup{colorlinks=true,allcolors=blue}

\begin{document}

\title{Water-Driven Porous Silicon as a Platform for Blue Materials}

\author{Marc Thelen}
\email{marc.thelen@tuhh.de}
\affiliation{Institute for Materials and X-ray Physics, Hamburg University of Technology, Hamburg, Germany}
\affiliation{Centre for X-ray and Nano Science CXNS, Deutsches Elektronen-Synchrotron DESY, Hamburg, Germany}
\author{Julia Brandt}
\affiliation{Institute of Photonics, Optics of Structured Media, Hamburg University of Technology, Hamburg, Germany}
\author{Stella Gries}
\affiliation{Institute for Materials and X-ray Physics, Hamburg University of Technology, Hamburg, Germany}
\affiliation{Centre for X-ray and Nano Science CXNS, Deutsches Elektronen-Synchrotron DESY, Hamburg, Germany}
\author{Manuel Brinker}
\affiliation{Institute for Materials and X-ray Physics, Hamburg University of Technology, Hamburg, Germany}
\affiliation{Centre for X-ray and Nano Science CXNS, Deutsches Elektronen-Synchrotron DESY, Hamburg, Germany}
\author{Alexander Petrov}
\affiliation{Institute of Photonics, Optics of Structured Media, Hamburg University of Technology, Hamburg, Germany}
\author{Patrick Huber}
\email{patrick.huber@tuhh.de}
\affiliation{Institute for Materials and X-ray Physics, Hamburg University of Technology, Hamburg, Germany}
\affiliation{Centre for X-ray and Nano Science CXNS, Deutsches Elektronen-Synchrotron DESY, Hamburg, Germany}
\affiliation{Centre for Molecular Water Science CMWS, Deutsches Elektronen-Synchrotron DESY, Hamburg, Germany}
\affiliation{United Nations University Hub on Engineering to Face Climate Change at the Hamburg University of Technology, United Nations University Institute of Water, Environment and Health, 21073 Hamburg, Germany}

\date{\today}

\begin{abstract}
Living organisms control water to transport matter, transfer energy and drive chemical reactions. In this perspective, we argue that porous silicon can give engineering materials comparable control over water. Its pore architecture, surface chemistry and semiconducting skeleton can be shaped across length scales with the precision of semiconductor processing and, like the water it can host, it relies on one of the most abundant chemical elements on Earth. Liquids in porous silicon, most often water, can flow as a continuum through pores a few nanometers wide, strain the skeleton through capillary and electrosorption stresses, shift structural color, switch transparency, and modulate or generate electrical signals. We describe three directions to combine and extend these effects and overcome their limitations. Pore hierarchy resolves the classical trade-off between interface area and transport speed by combining large and small pores and thereby allows functions to be scaled from thin planar films to three-dimensional components. Surface modification becomes a design variable that tunes the interaction between water and the pore wall. It can be set locally and, with switchable coatings, on demand. Finally, pores only a few molecules wide alter the properties of water itself. In the absence of a bulk-like volume, water can become dielectrically anisotropic and highly conductive. Altogether, these steps turn porous silicon from a responsive material into an adaptive, water-driven material.
\end{abstract}


\maketitle

\section{Introduction}

Water-driven materials make the state of confined water a design dimension in its own right. This is the central idea behind the Hamburg Cluster of Excellence BlueMat~\citep{BlueMat2026}. We study materials whose properties can be switched reversibly by pure water or by an aqueous working phase, so that the composition and state of solutions and electrolytes become design variables themselves. In this new class of engineering materials, which we call ``Blue Materials'', physical effects driven by water determine the function.

A host material for water-driven functions should combine a large accessible interface, controllable transport pathways, mechanical robustness and compatibility with established fabrication processes. Porous silicon, with its well-established etching techniques and well-known surface chemistry~\citep{Sailor2012,Canham2015}, is a natural candidate. With recent advances in fabrication and characterization, its interaction with water can be designed and monitored in considerable detail. Hierarchical porous silicon, for instance, can be fabricated at wafer scale~\citep{Gries2023}, water imbibition can be tracked optically on millisecond timescales~\citep{Dittrich2025Thesis} and the mechanical properties of the porous skeleton can be characterized without contact by laser ultrasonics~\citep{Thelen2021} or in operando by X-ray diffraction at synchrotron sources such as PETRA III~\citep{Brinker2022a}.

Throughout this perspective, we follow water through transport, mechanics, optics and electrical response toward adaptive, hierarchical materials and pores that hold only interfacial water, blurring the boundaries between the fields. We argue that confined water can turn porous silicon from a passive semiconductor host into an adaptive material and invite the porous silicon community to design for water.

\begin{figure*}[t]
\centering
\includegraphics[width=\textwidth]{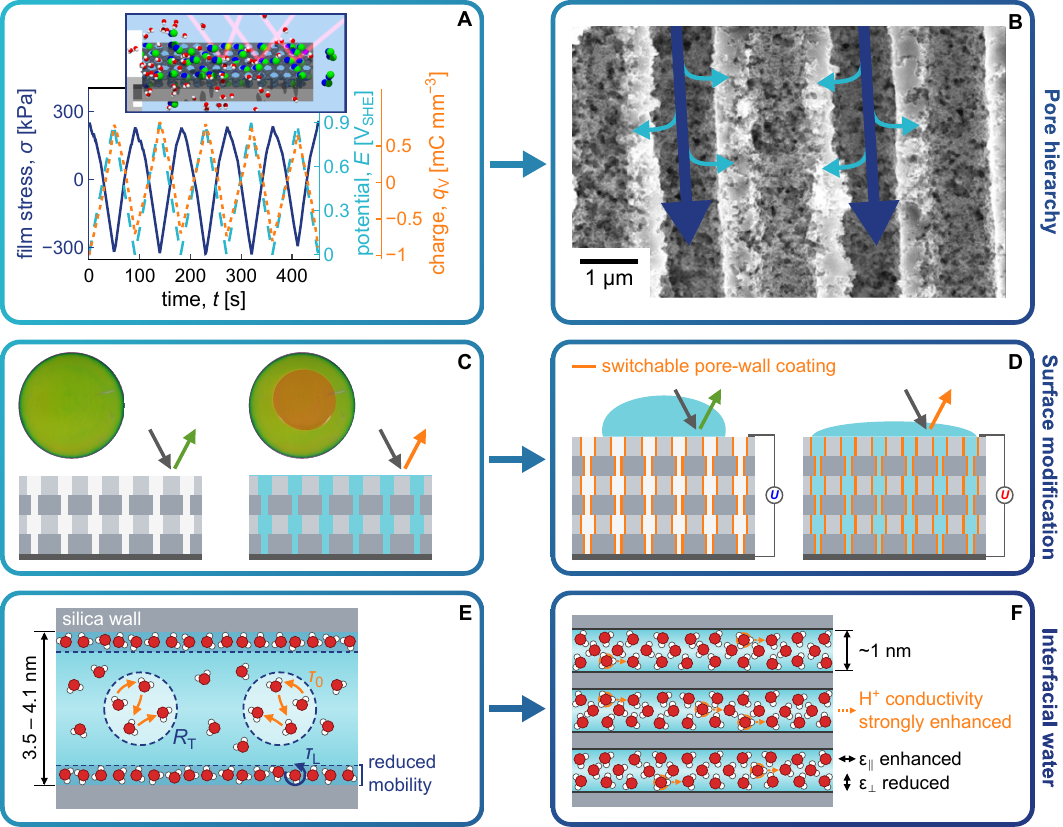}
\caption{Water-driven porous silicon today (left) and in perspective (right), shown for pore hierarchy, surface modification and interfacial water. \textbf{(A)}~Electrosorption actuation of a mesoporous-silicon cantilever in isotonic saline. Film stress $\sigma$, potential $E$ (vs.\ SHE) and volume-specific charge $q_\mathrm{V}$ over five potential cycles (data and cantilever rendering from \citet{Brinker2022}, CC~BY). \textbf{(B)}~In hierarchical porous silicon, macropores supply water (large arrows) and short paths lead into the mesoporous walls (small arrows), which could speed up actuation and other processes (SEM adapted from \citet{Gries2023}, CC~BY). \textbf{(C)}~Water red-shifts the stop band of a porous-silicon photonic crystal. The process is either passive or can be driven externally, e.g., by controlling the humidity. Photographs show a dry and a water-filled film. \textbf{(D)}~A switchable pore-wall coating makes the skeleton itself the active component and lets a drop bead up (off) or imbibe (on), thereby changing the reflected color. \textbf{(E)}~Water in 3.5--4.1~nm silica and organosilica mesopores. A reduced-mobility layer of 0.1--0.5~nm, jump diffusion (residence time $\tau_0$) within spheres of radius $R_\mathrm{T}$ and local relaxation ($\tau_\mathrm{L}$), drawn after \citet{Jani2021}. \textbf{(F)}~Oxidized porous silicon with $\sim$1~nm pores and a carbon lining (dark), holding only a few molecular layers of water. The dielectric constant of the interfacial water would become strongly anisotropic, enhanced along the pore ($\parallel$) and reduced perpendicular to it ($\perp$), as measured in graphite and boron nitride slits~\citep{Fumagalli2018,Wang2025} and simulated in carbon nanotubes~\citep{Loche2019}. In these slits, the conductivity along the slit is also strongly enhanced, which is attributed to proton transport (orange)~\citep{Wang2025}.}\label{fig:panels}
\end{figure*}

\section{Porous Silicon as a Platform for Water}

Porous silicon offers three main handles on water: pore architecture, wall chemistry and pore size. Its porosity follows the etching current density, so that modulating the current in time produces defined layer sequences~\citep{Pacholski2013}. As etched, the pores are aligned along the etching direction and randomly distributed in the plane~\citep{Sailor2012,Canham2015}. Combining photolithographically defined macropores with metal-assisted chemical etching~\citep{Huang2011,Kolasinski2014,Alhmoud2021} adds a second pore level, namely pores about 60~nm in diameter in the macropore walls~\citep{Gries2023}. Such hierarchical architectures assign transport and interface to different pore length scales, as the vascular systems of plants do~\citep{TyreeZimmermann2002}: large pores conduct water rapidly, whereas small pores provide a large internal surface and strong capillary forces.

Wall chemistry determines how water interacts with the pore walls. Etching in hydrofluoric acid, a reaction in which water takes part~\citep{Kolasinski2003,Kolasinski2009}, leaves hydrogen-terminated, hydrophobic walls~\citep{WuSailor2013}. In contrast, oxidized pore walls are hydrophilic~\citep{WuSailor2013} and pull water in spontaneously by imbibition and capillary condensation. Silica surfaces in contact with water can even drive redox chemistry~\citep{LiKolasinskiZare2023,Kolasinski2025}. Silanes can set the water contact angle between about 13$^\circ$ and 135$^\circ$~\citep{Velleman2010}. A thin carbon layer grown by pyrolysis of acetylene keeps the skeleton stable and electrochemically active in aqueous electrolyte~\citep{Salonen2004,Chen2011,GuoPrietoSimon2019,GuoPrietoSimon2022}. Liquid masking, i.e., temporarily filling the pores with an inert liquid, functionalizes the outer surface and the pore interior separately~\citep{WuSailor2013}, so that the placement of chemistry, not only its average, becomes a design variable.

Pore size, finally, sets how much of the water touches a wall. Etching reaches average pore sizes near 3~nm~\citep{Loni2015}. Atomic layer deposition narrows the pores further by coating their walls conformally~\citep{Elam2003,Puurunen2005}. Thermal oxidation narrows them through volume expansion and turns the skeleton into insulating silica~\citep{Canham2015}, whereas a carbon layer keeps the walls conductive. In short, architecture programs transport, wall chemistry determines the interaction with water and the pore size defines the state of the water itself.

\section{Water as a Transport Working Fluid}

Remarkably, water flow in mesopores is still well described by continuum hydrodynamics~\citep{Bocquet2010}. In porous silicon with 3~nm pores, permeation, imbibition and poroelastic relaxation follow Kelvin--Laplace thermodynamics and Darcy flow, with the remaining deviation interpreted as an immobile monolayer of water at the pore wall~\citep{Vincent2016}. Capillary rise of water in nanoporous glass points to an immobile layer of about 0.5~nm~\citep{Gruener2009Water} and quasielastic neutron scattering in 3.5--4.1~nm silica and organosilica pores finds a layer of 0.1--0.5~nm with strongly reduced mobility~\citep{Jani2021} (Figure~\ref{fig:panels}E). Simulations link such layers to the wall chemistry. At polar, hydrophilic surfaces, water molecules are transiently bound and the viscosity rises within a thin surface layer, whereas hydrophobic surfaces develop a depletion layer and finite slip~\citep{Sendner2009}. In films about a nanometer thick, these viscous interfacial layers dominate the hydration friction~\citep{Schlaich2017}.

Thin-film interference also lets porous silicon report its own filling~\citep{Talukdar2020}. Laser interferometry of capillary filling yields the hydraulic radius and tortuosity of the pore network~\citep{Acquaroli2011}. A photonic microcavity embedded in a layer resolves imbibition globally and locally in the same reflected spectrum. The complete imbibition event lasts only a few milliseconds, well within reach of spectrometers that acquire on microsecond timescales. Microcavities at several depths map water sorption through the film without sectioning it, revealing type-V water isotherms with hysteresis and capillary condensation above about 70\,\% relative humidity in surface-oxidized films~\citep{Dittrich2025Thesis}.

The transport mechanism is not fixed by the material alone. In mesoporous silicon with pores about 4~nm in diameter, the imposed humidity switches water uptake from a diffusion-like process to an imbibition-like front~\citep{Vincent2017}. In bilayers of large and small pores, the stacking sequence reverses the direction of the drying front and changes its rate, as shown with ethanol~\citep{Cencha2025Directional}. Drying is not simply filling in reverse, however. Microcavities at different depths in a porous-silicon film emptied without a resolvable delay and drying was much slower than estimated from capillary flow or Knudsen diffusion, which suggests percolation-controlled emptying~\citep{Dittrich2025Thesis}. Humidity and architecture thus determine not only how water enters, but also how it leaves.

A moving front does more than fill pores. It continuously creates fresh solid--liquid interface and with it the electrical double layers, interfacial stresses and optical contrast exploited below. Steering the front, actively or passively, would make this generation of interface controllable. In nanoporous gold, electrocapillarity switches the imbibition of an aqueous electrolyte against cyclohexane on and off within a potential window of less than 1~V~\citep{Xue2014}. To our knowledge, porous silicon has not been used to steer imbibition electrically, although its interface can be charged electrochemically~\citep{Brinker2022} and its fronts can be tracked optically. We therefore propose segmented electrodes that modulate the interfacial energy along a membrane while the front is followed in the reflected spectrum. Alternatively, tailored pore profiles could act passively as capillary valves or, with partially wetting walls, even as diodes that admit water from one side only. Janus membranes, whose two faces differ in wettability, already act as such diodes~\citep{ZhouGuo2019}. A wood membrane, for instance, transports water only from its hydrophobic to its hydrophilic face~\citep{DingKeplinger2020} and conical micropores can add a geometric asymmetry~\citep{DaiWang2019}. Porous silicon can provide both asymmetries, since liquid masking renders only the pore entrances hydrophobic~\citep{WuSailor2013} and the etching current sets the porosity profile through the thickness~\citep{Pacholski2013}. Light offers another route, e.g., by switching the water contact angle of spiropyran-coated porous silicon by about 11$^\circ$~\citep{Dattilo2007}. A projected light pattern could modulate wetting locally.

\section{Water as a Mechanical Working Fluid}

Water condensing in porous silicon strains the skeleton. In 3~nm pores the confined water sustains tensions of up to about 100~MPa~\citep{Vincent2016}. For a membrane traversed by channels 8~nm in diameter, the strain isotherm yields the pore-load modulus, which links the Laplace pressure of the condensed water to the deformation~\citep{Gor2015Capillary}. With organic adsorbates, X-ray diffraction resolves the contraction and expansion of the pore walls during sorption~\citep{Dolino1996} and the sorption strain is anisotropic~\citep{Grosman2015}. During water imbibition into mesoporous silica, the matrix expands with the square root of time instead of contracting and its length jumps when the menisci reach the top surface. \citet{Sanchez2024} attributed this to expansive surface stresses (Bangham effect) and to the build-up and release of Laplace pressures as the menisci advance, arrest or disappear. Simple deformation measurements can thus quantify surface stresses, Laplace pressures and transport.

Water can also act as the electrolyte that makes the mechanics electrically addressable. Without any functionalization, charging the internal interface of oxidized mesoporous silicon in aqueous perchloric acid or isotonic saline produces reversible film stresses of up to about 600~kPa for potential variations of about 1~V~\citep{Brinker2022}. This response results from the concerted action of roughly $10^{11}$ parallel pores per square centimeter. Because the pore walls are semiconducting, space charge contributes alongside the electrical double layer. In particular, the nanopores shift the flat-band potential relative to bulk silicon~\citep{Brinker2024MottSchottky}. Filled with polypyrrole, the pores host an actuating polymer phase that produces a macroscopic electrostrain in aqueous electrolyte~\citep{Brinker2020}. In operando X-ray diffraction resolves the resulting strain of the silicon pore walls~\citep{Brinker2022a}. Because humidity and potential act on the same interface, we envision actuators in which humidity sets the operating point while the potential trims it.

Converting interfacial stress into displacement requires the elastic constants of the porous layer, which laser-excited guided waves, including local zero-group-velocity (ZGV) resonances, determine without contact. For a free-standing membrane with 55\,\% porosity, analyzed under the assumption of transverse isotropy, the stiffness was reduced by about 80\,\% compared with bulk silicon, with longitudinal stiffness coefficients $C_{11}\approx22$~GPa in plane and $C_{33}\approx33.2$~GPa parallel to the pores~\citep{Thelen2021}.

Ultrasonic studies of water in Vycor and controlled-pore glass found the confined water stiffer than bulk water, with a bulk modulus that depends linearly on the Laplace pressure~\citep{Ogbebor2023}. Laser ultrasonics would allow such effects to be studied separately along and across the pores, with systematic control of the pore size and thus of the confinement strength.

\section{Water as an Optical Working Fluid}

A photonic crystal made from porous silicon layers of alternating porosity acts as a Bragg reflector or, with a defect layer, a microcavity~\citep{TorresCosta2010,Pacholski2013,Ivanov2020}. Water inside these pores increases the effective refractive index of each layer and, thus, red-shifts the photonic stop band (Figure~\ref{fig:panels}C), analogous to the way water uptake turns the scales of the longhorn beetle \textit{Tmesisternus isabellae} from gold to red through swelling of the multilayer and a higher effective refractive index~\citep{Liu2009}. An effective medium model translates the refractive-index change into a volume fraction of water. For light polarized perpendicular to cylindrical pores, the Maxwell Garnett model is accurate even at high porosity~\citep{Brandt2024}. Varying the hydrophilicity from layer to layer would enhance the refractive-index contrast during partial pore filling and thus widen the photonic band gap.

Other optical effects respond to pore filling as well. In porous silicon metasurfaces supporting Fano guided-mode resonances, liquid infiltration shifts the structural color and a water aerosol provides a visible image contrast~\citep{Dash2026}. Encoding local filling thresholds in a lateral variation of porosity or oxidation would let a wetting front write an image that is erased again during drying. Besides being visible to an observer, the color change could feed back into a valve or an actuator.

Beyond color, pore filling also modulates light scattering. A fully oxidized porous silicon membrane is transparent when saturated with water but turns opaque during drying, while wet and dry clusters coexist~\citep{Dittrich2025Thesis}. A mesoporous SiO$_2$ membrane shows the same behavior: water increases its visible transmittance from 43\,\% (dry) to 81\,\% (filled), whereas partial emptying lowers it to 3\,\%. This was traced back to scattering at discontinuous water--air interfaces~\citep{Guo2024}. A white membrane of oxidized hierarchical porous silicon likewise turns transparent when it takes up an index-matched liquid~\citep{Gries2023}. Such switching could serve as adaptive shading or privacy glazing. Controlling the water content locally could create images on an otherwise transparent porous membrane.

\section{Water as an Electrical Working Fluid}

Between 25 and 75\,\% relative humidity, the DC conductance of porous silicon rises by three orders of magnitude, with time constants of several days~\citep{Mares1995}. Porous-silicon humidity sensors exploit such changes, e.g., as capacitors with a porous-silicon dielectric, whose effective dielectric constant rises as water replaces air in the pores~\citep{Rittersma2000}, or through impedance measurements on oxidized layers~\citep{Kovacs2009}. Water can also generate electricity by intrusion and extrusion in hydrophobic nanoporous silicon monoliths, producing a triboelectric output~\citep{Bartolome2025}. Replacing H$_2$O by D$_2$O on a single such monolith revealed that immersion--emersion, droplet contact and pressure-driven intrusion--extrusion are distinct electrification regimes: the isotope substitution inverted the polarity upon immersion--emersion but preserved it upon intrusion, where the peak voltage rose from about 94~mV to 219~mV~\citep{Arkan2026}. The electrical response thus depends on whether water only wets the surface roughness or fills the pores.

Hydrovoltaic devices convert the evaporation of water into electricity. \citet{Xue2017} demonstrated that carbon black sheets deliver sustained voltages of up to 1~V and similar devices have been created with silicon nanowires~\citep{Qin2020,Shao2021SiliconHydrovoltaic,Shao2022SiNWHydrovoltaic}. Electrochemically etched porous silicon could add a semiconducting skeleton, a tunable interfacial charge and controllable evaporation fronts. The defensible next step is an experiment rather than a performance claim. Such an experiment could use a membrane with independently contacted ends, operated under controlled humidity, evaporation rate, ionic strength and flow direction. Reversing its hydraulic and geometric asymmetries and comparing blocking (purely capacitive) with non-blocking electrodes would separate streaming, diffusion, redox and contact-potential contributions. Recording the liquid distribution optically on the same membrane would relate the electrical signal to the local water state, a prerequisite for the adaptive behavior discussed below.

\section{Discussion: From Planar Films to Adaptive Silicon and Interfacial Water}\label{sec:ultra}

Nearly all demonstrations so far use planar films, membranes or cantilevers, ideal calibration elements with known transport length, optical response and conversion from surface stress to curvature (Figure~\ref{fig:panels}A). One central opportunity lies in three-dimensional components in which water access, conversion into a response and a reproducible reset are designed together. Hierarchical porous silicon already separates access channels from nanoscale internal area~\citep{Gries2023} (Figure~\ref{fig:panels}B). We are extending the concept by realizing three-dimensional lattices, in which struts or shells set the deformation mode, macropores deliver and vent water and mesoporous faces generate capillary or electrosorption stress while reporting the saturation through color. Even a single etched membrane should bend with humidity, as its cone-shaped pores create a porosity gradient~\citep{Thelen2021}; a mesoporous silica film on a silicon cantilever already bends largely reversibly with humidity~\citep{Ganser2016}. Since the outputs discussed here are usually measured on different samples and often with different fluids, a decisive experiment would treat opposite faces of one strut differently, by pore radius, oxidation or electrode potential and follow saturation, curvature, impedance, laser-ultrasonic stiffness and in situ X-ray scattering during filling and reset.

Surface chemistry decides the function of a given pore domain. A pore coating sets the wall chemistry once, whereas a switchable pore-wall coating adds a control variable (Figure~\ref{fig:panels}C,D). On a porous silicon Bragg stack, a water droplet would bead up while the coating is non-wetting and imbibe into the pores once switched to wetting, red-shifting the stop band. Combined with the electrochemically addressable skeleton~\citep{Brinker2022}, wetting could be written electrically and read optically.

Such components point from responsive materials, which follow their environment as stimuli-responsive polymers do~\citep{Stuart2010}, to adaptive ones, which change their response according to their history. Natural hierarchical materials store this history in their structure through growth and remodeling~\citep{Fratzl2007}. Nanofluidic memory has been shown in two-dimensional channels filled with aqueous electrolytes~\citep{Robin2023} and in a protein nanopore whose memristive conductance was traced back in simulations to hydrophobic wetting and dewetting~\citep{Paulo2023}. In porous silicon, information could be stored in the distribution, hydration state or ionic configuration of the confined water. Sorption hysteresis is another simple natural candidate for such a memory. The sequence write, retain, read and erase can be tested on a single sample with an optical or mechanical readout. Furthermore, pores behind narrow constrictions empty by pore blocking or by cavitation, depending on constriction size and temperature~\citep{Casanova2012,Bossert2021}. Water also drains in bursts by cavitation from micrometer-sized voids behind nanoscale throats~\citep{Vincent2014Drying}.

Interfacial water marks a further extension. Once a pore is only a few molecules wide, no bulk-like core remains and the material properties treated above as material constants are set by the interface alone. In graphite and boron nitride slits narrower than about 2~nm, the dielectric constant of water at low frequencies drops from 80 to 2 in the direction orthogonal to the slit~\citep{Fumagalli2018} but rises to about 1,000 parallel to it, with an in-plane conductivity, attributed to protons, comparable to Nafion~\citep{Wang2025}. Simulations predict this anisotropy to be more pronounced in cylindrical than in planar pores~\citep{Loche2019} (Figure~\ref{fig:panels}F). Anomalously fast proton conduction has also been observed in nanoporous matrices of diamond grains, with a conductivity up to five orders of magnitude above that of bulk water~\citep{Artemov2020}. In 1-nm clay channels that restrict all water to thin interfacial layers, the water conducts well enough to serve as the only electrolyte of a supercapacitor~\citep{Artemov2026}. Spectroscopy of electrolyte films between graphene and CaF$_2$ shows that their structure remains the sum of the two interfaces down to gaps of about 0.8~nm~\citep{WangTang2025}. Whether water at hydride- or silica-terminated silicon walls conducts likewise remains to be shown. In angstrom-scale slits, ions pass by deforming their hydration shells~\citep{Esfandiar2017}, a water monolayer excludes all ions but admits protons~\citep{Gopinadhan2019} and capillary condensation still follows the bulk Kelvin equation~\citep{YangSun2020}. Such atomically precise slits~\citep{Radha2016} are difficult to scale. The carbon lining introduced above could bring carbon walls to porous silicon, which is etched at wafer scale. In porous silicon, the solid side of the interface would change as well. Walls only a few nanometers thick widen the band gap, which lets porous silicon emit in the visible~\citep{Canham1990,Cullis1997}. Such walls are expected to have a reduced dielectric constant because of their surface bonds~\citep{Delerue2003}.

The regime could be reached by the etching and narrowing routes introduced above. Narrowing only the mesopores of a hierarchical membrane keeps the macropores as access channels, without which a membrane of such narrow pores would have no throughput. Several functions would then acquire distinct signatures. Optically, the water would carry a uniaxial anisotropy that is switched by filling and read out in the perpendicular polarization, for which the Maxwell Garnett model of cylindrical pores is validated~\citep{Brandt2024}. Electrically and mechanically, proton-selective conduction with steric ion exclusion and the pore-load modulus could be probed in the narrowest pores.

Decades of work on porous silicon have provided precise control over pore geometry and surface chemistry. Applying this control to water-driven functions is, in our view, the logical next step toward new functionality that does not depend on critical elements. Microelectronics followed a simple rule of thumb: if you can do it with silicon, do it with silicon. Water-driven materials suggest its natural extension: if you can do it with water, do it with water.

\section*{Conflict of Interest Statement}
The authors declare that the research was conducted in the absence of any commercial or financial relationships that could be construed as a potential conflict of interest.

\section*{Author Contributions}
MT: Conceptualization, Writing -- original draft, review \& editing. JB: Writing -- review \& editing. SG: Writing -- review \& editing. MB: Writing -- original draft, review \& editing. AP: Writing -- original draft, review \& editing. PH: Conceptualization, Supervision, Funding acquisition, Writing -- original draft, review \& editing.

\section*{Funding}
This work was supported by the Deutsche Forschungsgemeinschaft (DFG, German Research Foundation) through the Excellence Strategy of the Federal Government and the federal states -- EXC 3120/1 BlueMat: Water-Driven Materials (533771286), through the project ``Aqueous Electrolytes in Nanoporous Media: Structure, Dynamics and Electrochemo-Mechanical Actuation'' (Project number 509293944), and ''Dynamic Electrowetting at Nanoporous Surfaces: Switchable Spreading, Imbibition, and Elastocapillarity'', Project number 422879465 (SPP 2171).

\section*{Acknowledgments}
Claude Opus 5.5 (Anthropic PBC) was used for language and stylistic refinement. The authors rigorously reviewed the text and take full responsibility for the final publication.

\section*{Data Availability Statement}
No new datasets were generated or analyzed for this article. The data shown in Figure~\ref{fig:panels}A are taken from \citet{Brinker2022}.

\bibliography{references}

\end{document}